# STATUS AND PLANS FOR A SUPERCONDUCTING RF ACCELERATOR TEST FACILITY AT FERMILAB*

J. Leibfritz[†], R. Andrews, C. Baffes, K. Carlson, B. Chase, M. Church, E. Harms, A. Klebaner, M. Kucera, A. Martinez, S. Nagaitsev, L. Nobrega, P. Piot, J. Reid, M. Wendt, S. Wesseln, FNAL, Batavia, IL 60510, USA

## Abstract

The Advanced Superconducting Test Accelerator (ASTA) is being constructed at Fermilab. The existing New Muon Lab (NML) building is being converted for this facility. The accelerator will consist of an electron gun, injector, beam acceleration section consisting of 3 TTF-type or ILC-type cryomodules, multiple downstream beam lines for testing diagnostics and conducting various beam tests, and a high power beam dump. When completed, it is envisioned that this facility will initially be capable of generating a 750 MeV electron beam with ILC beam intensity. An expansion of this facility was recently completed that will provide the capability to upgrade the accelerator to a total beam energy of 1.5 GeV. Two new buildings were also constructed adjacent to the ASTA facility to house a new cryogenic plant and multiple superconducting RF (SRF) cryomodule test stands. In addition to testing accelerator components, this facility will be used to test RF power systems, instrumentation, and control systems for future SRF accelerators such as the ILC and Project-X. This paper describes the current status and overall plans for this facility.

## INTRODUCTION

Fermi National Accelerator Laboratory is constructing the Advanced Superconducting Test Accelerator (ASTA) as part of its new state-of-the-art superconducting radio frequency (SRF) test facility complex. The goal is to test SRF cryomodules with beam for the development of next generation high intensity linear accelerators, such as Project-X and the International Linear Collider (ILC).

Initially conceived as a test area for the ILC R&D program, the objective was to build a freestanding electron linac, capable of testing one RF Unit with ILC beam intensity [1]. An ILC RF Unit consists of three ILC-type cryomodules, each containing eight high gradient 9-cell 1.3 GHz niobium SRF cavities, powered by a single 10 MW pulsed RF system. The design beam parameters for this facility are shown in Table 1.

Table 1: Beam Parameters

| Bunch Charge | 3.2 nC |
|---|---|
| Pulse Length | 1 msec |
| Bunches Per Pulse | 3000 |
| Repetition Rate | 5 Hz |
| Bunch Length | < 300 µm RMS |
| Injector Beam Energy | 40 MeV |

| Final Beam Energy (1 RF Unit) | 750 MeV |
|---|---|

## ASTA FACILITY DESCRIPTION

ASTA is being built in the existing NML building and consists of a shielding cave which houses the accelerator, as well as areas for the support equipment and systems required for operation (Fig. 1). The accelerator consists of three distinct sections: the injector, SRF accelerator, and the test beam lines.

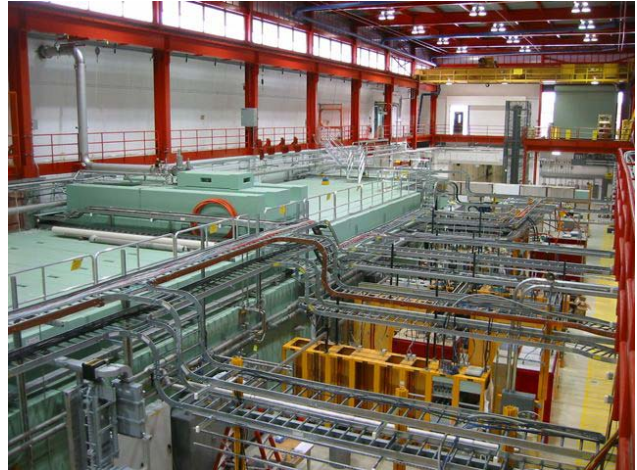

Figure 1: Advanced Superconducting Test Accelerator

The injector is comprised of a 1.3 GHz photocathode electron RF gun, which generates the electron beam [2], followed by two cryomodule cryostats called Capture Cavities I & II (CC1 & CC2). Each of these contains a single high-gradient 1.3 GHz 9-cell superconducting RF cavity. These devices provide the initial acceleration of the electrons to approximately 40 MeV.

Following the injector is the accelerator portion of ASTA, which consists of a series of 12-meter long SRF cryomodules. The first two cryomodules (CM1 and CM2) are a TTF Type-III+ design. The third cryomodule (CM3), which will complete the full RF Unit, will be an ILC-Type IV design [3]. It is expected that these three cryomodules will be capable of generating a beam energy of approximately 750 MeV.

Downstream of the accelerating cryomodules is the test beam line section, which consists of an array of multiple high-energy beam lines that transport the electron beam from the accelerating cryomodules to one of two beam absorbers. Each absorber is capable of dissipating up to 75 kW of power and is surrounded by a large steel and concrete shielding dump [4].

In addition to testing the accelerator components, the intent of this facility is to also test the support systems required for a future SRF linac. These systems include RF power, low-level RF, controls, instrumentation, low conductivity cooling water, and cryogenics.

## STATUS AND SCHEDULE

The construction of this facility is a large undertaking, which is being done in multiple phases. The first phase included the installation of the infrastructure necessary to operate the facility: cryogenics, water, power, RF, and controls, as well as building the test cave. Once the infrastructure was in place, CC2 (containing a single 9-cell cavity) and CM1 (containing eight 9-cell cavities) were installed, cooled down to 2 K (23.4 Torr), and RF power tested. This phase of the project recently concluded with the successful completion of the testing program of CM1 in March 2012, which yielded an overall accelerating gradient of approximately 200 MeV. CM1 is currently being replaced with CM2 (Fig. 2), which is anticipated to be the first high gradient cryomodule in which all of the cavities meet the ILC requirements of operating at 31.5 MeV/m. Cool down and initial operation of CM2 is expected in summer 2012.

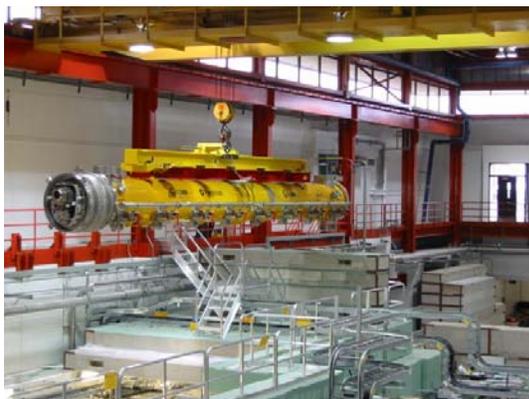

Figure 2: CM1 being removed from ASTA test cave

The next phase of the project involves preparing everything needed to generate the first beam in the accelerator. This includes: the installation and commissioning of the electron gun and associated laser system [2], which is expected to begin operation this summer; completion of the injector and test beam lines of the accelerator; and fabrication and installation of the high energy beam absorbers and dump, which was completed in December 2011 [4]. This phase is currently underway with first beam expected in 2012.

Following the generation of the first beam, the subsequent phases of the project will involve adding additional cryomodules to the accelerator section to complete the full RF Unit for testing.

## EXPANSION AND FUTURE PLANS

Two large construction projects were recently completed that involved expanding the length of the accelerator tunnel, upgrading the cryogenic system, and providing an area to test various types of cryomodules (Fig. 3). In addition to the SRF cryomodule tests, there are also plans for conducting advanced accelerator R&D (AARD) experiments at low energy (40-50 MeV) using a beam line that branches off of the injector, and also at high energy (250 MeV to 1.5 GeV) using the test beam lines. Additionally, there are now plans to also build a proton linear accelerator that will be used as a prototype for the front-end of Project-X.

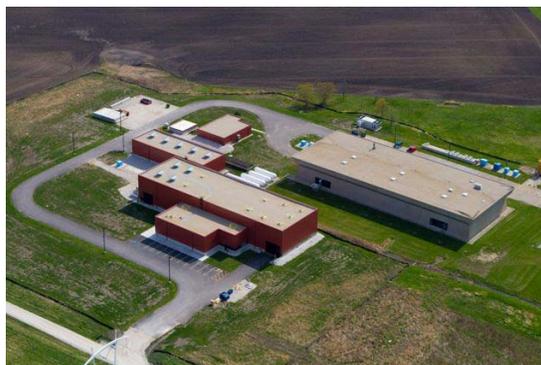

Figure 3: Newly constructed buildings at SRF Test Facility (existing NML building is on the right)

### NML Facility Expansion

The first construction project involved the addition of a 70 m-long tunnel to the North of the existing NML building, which houses ASTA. This expansion essentially doubled the length of the test accelerator from 75 m to 140 m. This will provide enough space for up to six (12 m-long) cryomodules and increase the beam energy capability of the facility from 750 MeV to 1.5 GeV.

In addition to providing additional length to the overall accelerator, the expansion also includes a large 15 m-wide area for the high-energy test beam lines at the downstream end of the accelerator. This space is intended to be a user's facility to carry out Fermilab's AARD program. Several AARD experiments have already been identified for ASTA, the first of which will be the Integrable Optics Test Accelerator (IOTA). IOTA is a 30 m-circumference storage ring that will be used to study non-linear accelerator optics [5].

An enclosure to house the high-energy beam absorbers and dump is situated at the end of the test beam lines and contains a pass-thru beam line that will allow for further expansion of the accelerator in the future.

### Cryomodule Test Facility

Two adjoining buildings called the Cryomodule Test Facility (CMTF) were constructed adjacent to the existing NML building (Fig. 3 and 4). Completed in December 2011, CMTF was designed to house a new cryogenic plant and stand-alone SRF cryomodule test stands. The smaller of the two building will house the noisy vibrating equipment (compressors, pumps, etc.) needed to operate the cryogenic plant. The other building will contain the cryogenic cold boxes, cryomodule test stands, RF systems, a vacuum clean room, and office area.

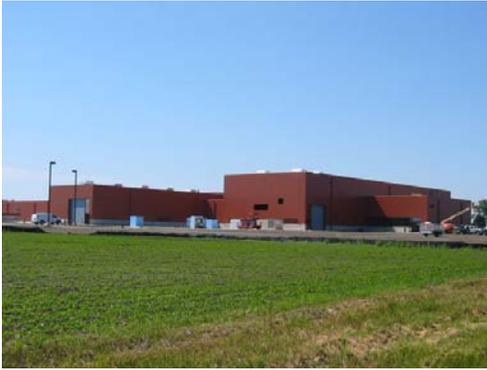

Figure 4: New CMTF buildings

The new cryogenic plant is designed to have the capability of supporting simultaneous operation of ASTA as well as the cryomodule test stands located in CMTF [6]. The original plan was for CMTF to house two test stands that were capable of testing various styles of cryomodules at 325 MHz, 650 MHz and 1.3 GHz, in pulsed and continuous wave (CW) modes of operation. However, in recent months, the plan has been modified to replace one of the cryomodule test stands with the Project-X Injector Experiment (PXIE) [7], an $H^-$ linear accelerator that will serve as a prototype for the front-end of Project-X. A layout of the entire SRF Accelerator Test Facility complex is shown in Figure 5.

## ACKNOWLEDGEMENTS

Authors would like to acknowledge technical specialists C. Exline, D. Franck, W. Johnson, R. Kellett and C. Rogers for all their efforts, as well as the entire NML Project Team.

## REFERENCES

[1] M. Church and S. Nagaitsev, "Plans for a 750 MeV Electron Beam Test Facility at Fermilab," PAC'07, Albuquerque, New Mexico, June 2007.

[2] S. Nagaitsev et al., "Fermilab L-band Electron Gun for the ILC Cryomodule Test Facility," PAC'07, Albuquerque, New Mexico, June 2007.

[3] T. Arkan et al., "Superconducting RF Cryomodule Production & Testing at Fermilab," LINAC'10, Tsukuba, Japan, September 2010.

[4] C. Baffes et al., "Mechanical Design of a High Energy Beam Absorber for the Advanced Superconducting Test Accelerator (ASTA) at Fermilab," these proceedings.

[5] S. Nagaitsev et al., "Design and Simulation of IOTA- a Novel Concept of Integrable Optics Test Accelerator," these proceedings.

[6] A. Martinez et al., "Design and Testing of the New

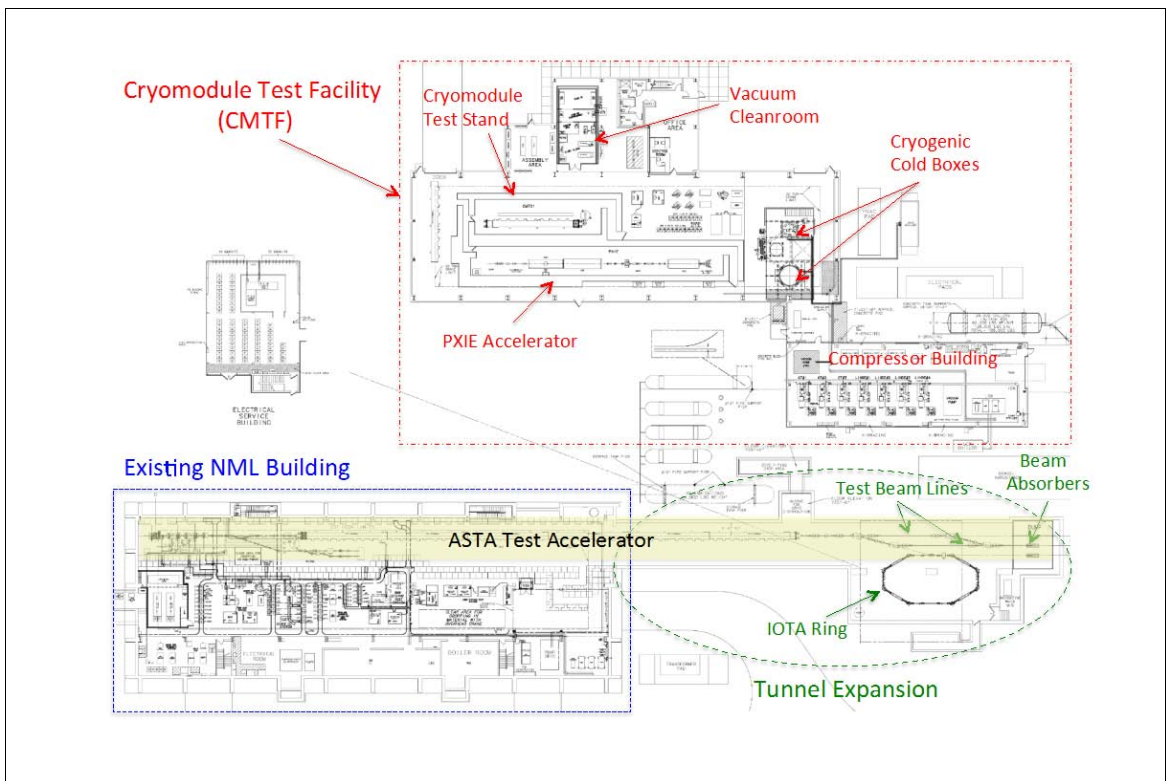

Figure 5:  Layout of SRF Accelerator Test Facility

Muon Lab Cryogenic System at Fermilab," CEC/ICMC 2009, Tucson, Arizona, June 2009.

[7] V. Lebedev et al., "PXIE: Project X Injector Experiment," these proceedings.